
\def\raggedcenter{\leftskip=4em plus 12em \rightskip=\leftskip
  \parindent=0pt \parfillskip=0pt \spaceskip=.3333em \xspaceskip=.5em
  \pretolerance=9999 \tolerance=9999
  \hyphenpenalty=9999 \exhyphenpenalty=9999 }
\def\refis{\item}
\font\twelverm=cmr10 scaled 1200    \font\twelvei=cmmi10 scaled 1200
\font\twelvesy=cmsy10 scaled 1200   \font\twelveex=cmex10 scaled 1200
\font\twelvebf=cmbx10 scaled 1200   \font\twelvesl=cmsl10 scaled 1200
\font\twelvett=cmtt10 scaled 1200   \font\twelveit=cmti10 scaled 1200

\skewchar\twelvei='177   \skewchar\twelvesy='60


\def\twelvepoint{\normalbaselineskip=12.4pt
  \abovedisplayskip 12.4pt plus 3pt minus 9pt
  \belowdisplayskip 12.4pt plus 3pt minus 9pt
  \abovedisplayshortskip 0pt plus 3pt
  \belowdisplayshortskip 7.2pt plus 3pt minus 4pt
  \smallskipamount=3.6pt plus1.2pt minus1.2pt
  \medskipamount=7.2pt plus2.4pt minus2.4pt
  \bigskipamount=14.4pt plus4.8pt minus4.8pt
  \def\rm{\fam0\twelverm}          \def\it{\fam\itfam\twelveit}%
  \def\sl{\fam\slfam\twelvesl}     \def\bf{\fam\bffam\twelvebf}%
  \def\mit{\fam 1}                 \def\cal{\fam 2}%
  \def\tt{\twelvett}
  \textfont0=\twelverm   \scriptfont0=\tenrm   \scriptscriptfont0=\sevenrm
  \textfont1=\twelvei    \scriptfont1=\teni    \scriptscriptfont1=\seveni
  \textfont2=\twelvesy   \scriptfont2=\tensy   \scriptscriptfont2=\sevensy
  \textfont3=\twelveex   \scriptfont3=\twelveex  \scriptscriptfont3=\twelveex
  \textfont\itfam=\twelveit
  \textfont\slfam=\twelvesl
  \textfont\bffam=\twelvebf \scriptfont\bffam=\tenbf
  \scriptscriptfont\bffam=\sevenbf
  \normalbaselines\rm}

\def\tenpoint{\normalbaselineskip=12pt
  \abovedisplayskip 12pt plus 3pt minus 9pt
  \belowdisplayskip 12pt plus 3pt minus 9pt
  \abovedisplayshortskip 0pt plus 3pt
  \belowdisplayshortskip 7pt plus 3pt minus 4pt
  \smallskipamount=3pt plus1pt minus1pt
  \medskipamount=6pt plus2pt minus2pt
  \bigskipamount=12pt plus4pt minus4pt
  \def\rm{\fam0\tenrm}          \def\it{\fam\itfam\tenit}%
  \def\sl{\fam\slfam\tensl}     \def\bf{\fam\bffam\tenbf}%
  \def\smc{\tensmc}             \def\mit{\fam 1}%
  \def\cal{\fam 2}%
  \textfont0=\tenrm   \scriptfont0=\sevenrm   \scriptscriptfont0=\fiverm
  \textfont1=\teni    \scriptfont1=\seveni    \scriptscriptfont1=\fivei
  \textfont2=\tensy   \scriptfont2=\sevensy   \scriptscriptfont2=\fivesy
  \textfont3=\tenex   \scriptfont3=\tenex     \scriptscriptfont3=\tenex
  \textfont\itfam=\tenit
  \textfont\slfam=\tensl
  \textfont\bffam=\tenbf \scriptfont\bffam=\sevenbf
  \scriptscriptfont\bffam=\fivebf
  \normalbaselines\rm}

\def\beginparmode{\endmode
  \begingroup \def\endmode{\par\endgroup}}
\let\endmode=\par
{\obeylines\gdef\
{}}
\def\singlespace{\baselineskip=\normalbaselineskip}
\def\oneandathirdspace{\baselineskip=\normalbaselineskip
  \multiply\baselineskip by 4 \divide\baselineskip by 3}

\def\doublespace{\baselineskip=\normalbaselineskip \multiply\baselineskip by 2}

\newcount\firstpageno
\firstpageno=2

\hsize=6.5truein
\vsize=8.9truein
\parskip=\medskipamount
\twelvepoint            
\doublespace            

\def\subhead#1{                 
  \vskip 0.25truein             
  {\raggedcenter #1 \par}
   \nobreak\vskip 0.25truein\nobreak}

\newcount\notenumber
\def\clearnotenumber{\notenumber=0}
\def\note{\advance\notenumber by1 \footnote{$^{\the\notenumber}$}}
\clearnotenumber
\font\bfone=cmbx10 scaled\magstep1
\font\cs=cmcsc10 scaled\magstep1
\font\itm=cmti10 scaled\magstephalf
\font\rrm=amr10 scaled\magstephalf
\font\sseight=cmssq8 scaled\magstep1

\def\bx{{\bf x}}

\def\rr{{\rlap{\sseight R} \, \hbox{\sseight R}}}
\topinsert\vskip 1 true in\endinsert
\beginparmode\doublespace
\centerline{\bfone CANONICAL QUANTIZATION AND BRAID INVARIANCE OF}
\smallskip
\centerline{{\bfone (2+1)-DIMENSIONAL GRAVITY COUPLED TO POINT
PARTICLES}\footnote{*}
{\tenrm This work is supported in
part by funds provided by the U. S. Department
of Energy (D.O.E.) under
contract \#DE-AC02-76ER03069, and in part by the Texas
National Research Laboratory Commission under grant \#RGFY92C6.}}
\beginparmode
\vskip 36pt
\singlespace
\centerline{\cs Daniel Kabat and Miguel E. Ortiz}
\vskip 3pt
\centerline{\itm Center for Theoretical Physics}
\centerline{\itm Laboratory for Nuclear Science}
\centerline{\itm and Department of Physics}
\centerline{\itm Massachusetts Institute of Technology}
\centerline{\itm Cambridge, Massachusetts\ \ 02139\ \ \ U.S.A.}
\centerline{\itm e-mails: kabat and ortiz @mitlns.mit.edu}
\vskip 0.7truein
\medskip
\medskip
\medskip
\vskip 0.7truein
\beginparmode\oneandathirdspace
{\rrm
\noindent
{\bf Abstract:}
\noindent
We investigate the canonical quantization of gravity coupled to pointlike
matter in 2+1 dimensions.  Starting from the usual point particle action in
the first order formalism, we introduce auxiliary variables which make
the action locally Poincar\'e invariant.  A Hamiltonian analysis shows that the
gauge group is actually larger than the Poincar\'e group -- certain additional
gauge constraints are present which act on the matter degrees of freedom.
These additional constraints are necessary to mimic the diffeomorphism
invariance present if the theory is formulated with a spacetime metric.  The
additional gauge constraints are realized projectively in the quantum theory,
with a phase in the composition law for finite gauge transformations.  That
phase is responsible for the braid invariance of physical observables
(holonomies).}
\vfill
\vskip -12pt
\noindent CTP \# 2210 \hfill  May 1993\par
\noindent hep-th/9305155\hfill
\vfill\eject\beginparmode

General relativity in 2+1 dimensions, in the absence of matter,
admits only flat geometries as classical solutions [1].  As a
result of this simplification the vacuum theory can be quantized in a
variety of ways [2]. Perhaps the simplest of these makes use of the
equivalence between 2+1 dimensional gravity and a Chern-Simons gauge
theory of the three dimensional Poincar\'e group ISO(2,1) [3,4].
However, as soon as matter is introduced the local Poincar\'e invariance
is broken and this approach fails.

Incorporating matter into a quantum theory of gravity in 2+1
dimensions seems to be extremely hard in all approaches.
It has so far only proven possible to discuss the
quantization of gravity coupled to point particles.  While there has been
much progress in understanding this problem [5,6],
a systematic quantization starting from the familiar action for a point
particle coupled to gravity has not been found.

In this letter we show that it is possible to perform a complete canonical
quantization, by first restoring the ISO(2,1) gauge symmetry in a way
proposed by Grignani and Nardelli [7]. We derive a description of
the quantum theory closely related to that given by Carlip [6],
where the non-trivial features of the quantization are contained in a
braiding condition on the wave function. Here, this condition is not imposed
after quantization, as in Ref. 6, but rather appears as a
consequence of a new constraint which generates translations of the
particles in the ambient spacetime.

The classical action for a spinless point particle coupled to gravity
in 2+1 dimensions may be written in first order form in terms of a dreibein
$e^a_\mu$ and a spin connection
$\omega^{ab}_\mu \equiv - \epsilon^{abc}\omega_{\mu
\,c}$\note{\tenpoint\noindent Greek
letters are spacetime indices.  Latin letters from the middle
of the alphabet $i,\,j,\,k$ are space indices, and latin letters from the
beginning of the alphabet $a,\,b,\,c$ are dreibein indices.  We take
$\eta_{ab} = {\rm diag}(1,\,-1,\,-1)$, $\epsilon^{012} = \epsilon^{12} = 1$,
and set $16 \pi G = 1$.}, as
$$
\eqalign{
S = S_{EH}+S_{PP}
=&- \int d^3x \, \epsilon^{\alpha\beta\gamma}  e_{\alpha a}
\left(\partial_\beta \omega_\gamma^a - \partial_\gamma \omega_\beta^a
+ \epsilon^a{}_{bc}\omega_\beta^b \omega_\gamma^c\right) \cr
&+ \int d\tau \,
\left[p_a
e^a_\mu(x(\tau)) {dx^\mu \over d\tau} - \lambda \left(\eta^{ab} p_a p_b -
m^2\right)\right].\cr}
\eqno(1)
$$
We shall work throughout this letter on a spacetime
with topology ${\rr}^3$.

To make the ISO(2,1) gauge symmetry of the gravitational part of the action
manifest, introduce a gauge field $A_\mu = e^a_\mu {\cal P}_a + \omega^a_\mu
{\cal J}_a$, where
the Poincar\'e generators ${\cal P}_a$, ${\cal J}_a$ obey the ISO(2,1)
algebra
$$
\lbrack {\cal P}_a, {\cal P}_b\rbrack = 0
\quad
\lbrack {\cal P}_a,
{\cal J}_b\rbrack = \epsilon_{abc} {\cal P}^c\quad
\lbrack {\cal J}_a,
{\cal J}_b\rbrack = \epsilon_{abc} {\cal J}^c.
$$
An invariant inner product on this algebra is given by $\langle {\cal J}_a,
{\cal P}_b\rangle=\eta_{ab}$, $\langle{\cal P}_a,{\cal P}_b\rangle =
\langle{\cal J}_a, {\cal J}_b\rangle = 0$.
Under an infinitesimal gauge transformation, $A_\mu \rightarrow A_\mu - D_\mu
\lambda$, $D_\mu \equiv \partial_\mu + \lbrack A_\mu, \, \cdot\, \rbrack$,
where $\lambda = \sigma^a {\cal P}_a + \tau^a {\cal J}_a$. The gravity part
of the action is manifestly invariant under this transformation since one
recognizes that $S_{EH}$ is the Chern-Simons action for the field $A_\mu$
[3,4].
The change in the point particle part of the action due to
the rotation $\tau^a$ can be absorbed by taking the momentum $p_a$ to likewise
transform with a local rotation, $p_a \rightarrow p_a - \epsilon_{abc} p^b
\tau^c$.  However no transformation of $p_a$ can absorb the change due to the
translation $\sigma^a$. It is in this sense that
invariance under local translations is broken by the matter coupling.

One may follow a procedure advocated by Grignani and Nardelli, and restore
local Poincar\'e invariance by gauging translations [7]. Introduce a
field $q^a(x)$ which transforms in the defining non-linear representation of
the Poincar\'e group, $q^a\rightarrow q^a-\epsilon^a{}_{bc} \tau^c+
\sigma^a$, and an ISO(2,1) covariant derivative
$D_\mu q^a = \partial_\mu q^a + \epsilon^a{}_{bc} \omega_\mu^b q^c +
e^a_\mu$.  $D_\mu q^a$ transforms covariantly
under local Poincar\'e transformations, i.e.~in the same
way that $\partial_\mu q^a$ behaves under global transformations:
$D_\mu q^a(x) \rightarrow D_\mu q^a(x) - \epsilon^a{}_{bc} D_\mu q^b(x)
\tau^c(x)$.  Since $D_\mu q^a$ transforms only with a rotation,
replacing $e^a_\mu$ with $D_\mu q^a$ everywhere in (1) makes the action
ISO(2,1) gauge invariant.  This gauging procedure is
available in any number of dimensions, but the result is simpler in 2+1 because
the $q^a(x)$ field drops out of the gravity part of the action. That $q^a$
drops out is not surprising, as the gravity action was already locally ISO(2,1)
invariant and didn't need to be gauged; formally it is a consequence of the
Bianchi identities [7].

A few remarks about the addition of the field $q^a(x)$:
\item{(a)}~The
introduction of the additional $q^a(x)$ degrees of freedom does not affect
the dynamics of the model.  As is manifest from the way they were
introduced, the $q^a(x)$ are gauge variant degrees of freedom, and one may
adopt the gauge condition $q^a(x) = 0$ to return to the original theory.
However, we shall see below that the constraints are
greatly simplified when the ISO(2,1) symmetry holds.
\item{(b)}~As the ISO(2,1)
invariance of the original gravity action was broken only at the location
of the particle, the field $q^a(x)$ only enters along the worldline as
$q^a(x(\tau))$. We define $q^a(\tau)\equiv q^a(x(\tau))$.
\item{(c)}~If more than one
particle is present, we introduce $q^a_A\equiv q^a(x_A(\tau))$ (where
$A=1,2,\ldots$ labels the various particles). We are able to quantize only by
treating the $q^a_A$ as independent variables. For consistency,
we must exclude coincident points $x_A=x_B$, ${A\ne B}$
from the configuration space.

We proceed with a discussion of the two particle case, although the
generalization to higher numbers of particles is straightforward.
To simplify the canonical structure, we
introduce a momentum $\pi_\mu^A$ conjugate to the position variable $x^\mu_A$,
with a Lagrange multiplier $u^\mu_A$ to constrain $\pi_\mu^A =
e^a_\mu(x_A(\tau)) p_a^A + \omega^a_\mu(x_A(\tau)) j_a^A$.  We also
gauge fix the reparameterization invariance of $\tau$, by setting
$x^0_A(\tau) = \tau$ and $\pi_0^A = e^a_0 p^A_a + \omega^a_0 j^A_a$.
Thus (1), after the replacement $e^a_\mu \rightarrow D_\mu q^a$, becomes
$$\eqalign{
S &= S_{EH} + \int dt\,\Bigl[ \pi_i^A { dx^i_A \over dt}
+ p_a^A {d q^a_A \over dt} - \lambda_A \left(\eta^{ab} p_a^A p_b^A
- m_A^2 \right)\cr
&
\qquad\qquad - u^i_A \left(\pi_i^A - e^a_i\left(t,\bx_A(t)\right)p_a^A
- \omega^a_i\left(t,\bx_A(t)\right)j_a^A \right)\Bigr]\cr }
\eqno (2)
$$
where $A=1,2$ labels the two particles, and
$j^a_A\equiv\epsilon^{abc}q^A_bp^A_c$.

Separating out the time derivatives in (2)
allows us to identify the canonical variables and their
Poisson brackets
$$\eqalign{
\lbrace p_a^A,\,q^b_B \rbrace &= \delta^b_a \delta^A_B\cr
\lbrace \pi_i^A,\,x^j_B \rbrace &= \delta^j_i \delta^A_B\cr
\lbrace e^a_i({\bf x}),\omega^b_j({\bf x}') \rbrace &= {1 \over 2} \eta^{ab}
\epsilon_{ij} \delta^2({\bf x} - {\bf x}')\cr}\eqno(3)$$
The variables $e_{0a}({\bf x})$, $\omega_{0a}({\bf x})$, $u^i_A$, and
$\lambda_A$ are Lagrange multipliers enforcing the constraints,
$$\eqalign{
\phi_R^a(\bx) &\equiv \epsilon^{ij} R^a_{ij}(\bx) - p_1^a \delta^2({\bf x} -
{\bf x}_1) - p_2^a \delta^2({\bf x} - {\bf x}_2) = 0\cr
\phi_T^a(\bx) &\equiv \epsilon^{ij} T^a_{ij}(\bx) - j_1^a \delta^2({\bf x} -
{\bf x}_1) - j_2^a \delta^2({\bf x} - {\bf x}_2) = 0\cr
\phi_{\pi\, i}^A &\equiv \pi_i^A - e^a_i({\bf x}_A)
p_a^A - \omega^a_i({\bf x}_A)
j_a^A = 0\cr
\phi_{p^2}^A &\equiv p_A^2 - m_A^2 = 0\cr}\eqno(4)
$$
where the ${\cal J}_a$ and ${\cal P}_a$ components of the ISO(2,1) field
strength are\note{\tenpoint\noindent A
torsion tensor may be constructed which, unlike
the field strength $T^a_{\mu\nu}$, vanishes on solutions to the
constraints [7].}
$$
\eqalign{
R^a_{\mu\nu} &= \partial_\mu \omega^a_\nu - \partial_\nu \omega^a_\mu
+ \epsilon^a{}_{bc} \omega_\mu^b \omega_\nu^c\cr
T^a_{\mu\nu} &= \partial_\mu e^a_\nu - \partial_\nu e^a_\mu +
\epsilon^a{}_{bc} \left(e_\mu^b \omega^c_\nu - e_\nu^b
\omega^c_\mu\right).\cr}
$$

The constraints have a straightforward interpretation.  The $\phi_R^a$
and $\phi_T^a$ constraints fix the curvature of the ISO(2,1) gauge field
and generate infinitesimal ISO(2,1) gauge transformations.
The $\phi_{p^2}^A$ constraints put the particles on-shell, and generate
the transformation $\lbrace \phi_{p^2}^A, q^a_B\} = 2 p^a_A \delta^A_B$,
but leave $p^a_A$ and $j^a_A$ invariant.  Finally, the
$\phi_{\pi\, i}^A$ constraints fix the momentum conjugate to $x^i_A$, and
generate infinitesimal translations of the particle coordinates:
$\lbrace \phi_{\pi \, i}^A , x^j_A \rbrace = \delta_i^j$.
Note that there are more constraints present than if this were just an
ISO(2,1) gauge theory.  This is expected because, had we formulated the model
with a spacetime metric, we would encounter diffeomorphism invariance.
Diffeomorphisms act by moving particles around, and cannot be fully represented
by the (ultralocal) action of the constraints of the ISO(2,1) gauge theory.

The algebra of the constraints is as follows: $\phi^a_{R}$ and $\phi^a_{T}$
reproduce ISO(2,1) as expected; the $\phi^A_{\pi\,i}$ have the only other
non-zero brackets,
$$\eqalign{
&\lbrace \phi_{\pi\, i}^1, \phi_{\pi\, j}^1 \rbrace
= -{1 \over 2} \epsilon_{ij} \left(\phi_T^a(\bx_1) p_a^1 + \phi_R^a(\bx_1)
j_a^1 \right) - {1 \over 2} \epsilon_{ij} \left( p_1 \cdot j_2 + p_2
\cdot j_1 \right) \delta^2\left({\bf x}_1 - {\bf x}_2 \right)\cr
&\lbrace \phi_{\pi\, i}^1, \phi_{\pi\, j}^2 \rbrace
= + {1 \over 2} \epsilon_{ij} \left( p_1 \cdot j_2 + p_2 \cdot j_1 \right)
\delta^2\left({\bf x}_1 - {\bf x}_2 \right)\cr
&\lbrace \phi_{\pi\, i}^2, \phi_{\pi\, j}^2 \rbrace
= -{1 \over 2} \epsilon_{ij} \left(\phi_T^a(\bx_2) p_a^2 + \phi_R^a(\bx_2)
j_a^2 \right) - {1 \over 2} \epsilon_{ij} \left( p_1 \cdot j_2 + p_2
\cdot j_1 \right) \delta^2\left({\bf x}_1 - {\bf x}_2 \right).\cr}
$$
This algebra closes on the physical configuration space with the coincident
point ${\bf x}_1={\bf x}_2$ excluded.

The constraint $\phi^A_{\pi\,i}$ is the mechanical momentum, or gauge
invariant velocity operator [8], of a particle moving in the gauge
connection $A^a_\mu$.  This suggests a useful analogy, namely the quantum
mechanics of a non-relativistic particle moving in a magnetic field.  In
that example, translations are generated by an abelian version of
$\phi^A_{\pi\,i}$, and these act non-trivially in a global sense, leading
to the derivation of the Dirac quantization condition for magnetic charge
[8].

Motivated by that abelian analysis, we proceed to quantize, adopting the
Poisson bracket algebra (3) for equal time commutators.  The constraints
(4) may be promoted to Hermitian operators; no operator ordering
difficulties arise. Locally, it is straightforward to see how the
constraints restrict a wave functional. The ISO(2,1) constraints restrict
the functional to be gauge invariant, and to have support only on ISO(2,1)
connections that are flat away from the sources, subject to a choice of
polarization [4]. The $\phi^A_{\pi\,i}$ constraints act
locally to remove any dependence of the functional on $x^i$ (we ignore the
reparameterization constraint for the moment). However, as we shall now
see, finite transformations generated by the $\phi^A_{\pi\,i}$ constraints
impose further conditions on the wave function.

By exponentiating the constraint $\phi_\pi^1$ we obtain a unitary operator
which performs a finite gauge transformation.
$$
U(a^i) = e^{-ia^i \phi_{\pi \, i}^1}
$$
As expected, $U(a^i)$ acts on states by translating particle 1 from ${\bf
x}_1$ to $\bx_1 + {\bf a}$.  It also parallel transports the ISO(2,1) charge of
particle 1 along a straight line from ${\bf x}_1$ to $\bx_1 + {\bf a}$, and it
changes the gauge field only along that line -- it leaves behind a (gauge
variant) string singularity in the Poincar\'e gauge field connecting the
initial and final locations of the particle:
$$\eqalign{
&U^\dagger({\bf a}) \bigl( e^a_i(\bx){\cal P}_a + \omega^a_i(\bx){\cal J}_a
\bigr) U({\bf a}) = e^a_i(\bx){\cal P}_a + \omega^a_i(\bx){\cal J}_a\cr
&\hbox to 2 true cm{} + {1 \over 2}\epsilon_{ij} \, a^j \int_0^1 d\tau \,
\delta^2(\bx_1 + {\bf a} \tau - \bx) {\rm P} \exp\Bigl\lbrace - \int_{\bx_1}
^{\bx_1 + {\bf a} \tau}(e + \omega)\Bigr\rbrace\bigl(p_1^a {\cal J}_a +
j_1^a {\cal P}_a\bigr).\cr}\eqno(5)$$
The parallel transport is along a straight path from $\bx_1$ to $\bx_1 +
{\bf a}\tau$.

One might expect that the operator $U({\bf a})$ should obey the abelian group
composition law for translations, $U({\bf a}_2) U({\bf a}_1) = U({\bf a}_1
+ {\bf a}_2)$.  Instead, explicit computation shows that a phase
(a 2--cocycle [8])
arises in the composition law, so that translations are realized
projectively,
$$
B \equiv U^\dagger(a_1^i + a_2^i) U(a_2^i) U(a_1^i).\eqno(6)
$$
The 2--cocycle measures the non-abelian flux of the ISO(2,1) gauge field,
pointing in the direction covariantly along $p_1^a {\cal J}_a + j_1^a
{\cal P}_a$, through a triangle with vertices located
at ${\bf x}_1$, ${\bf x}_1 + {\bf a}_1$ and ${\bf x}_1 + {\bf a}_1 +
{\bf a}_2$ (see Fig.~1):
$$\eqalign{
B &= {\rm P}_\tau \, \exp \Biggl[ i
\int_0^1 d\tau\, \int_0^\tau d\tau'\, a_1^i a_2^j \cr
&\qquad\qquad\quad \Biggl\langle
R^a_{ij}(\bx_1 + {\bf a}_1 \tau + {\bf a}_2 \tau')
{\cal J}_a + T^a_{ij}(\bx_1 + {\bf a}_1 \tau + {\bf a}_2\tau') {\cal P}_a,\cr
&\qquad\qquad\qquad {\rm P} \exp\Biggl\lbrace - \int_{\bx_1}^{\bx_1 + {\bf a}_1
\tau + {\bf a}_2 \tau'}(e + \omega)\Biggr\rbrace (p_1^a {\cal J}_a + j_1^a
{\cal P}_a)\Biggr\rangle\Biggr]\cr}$$
The cocycle vanishes unless particle 2 sits inside the triangle.
The symbol ${\rm P}_\tau$ denotes path ordering with respect to $\tau$ only;
the $\tau'$ integral sits inside the integrand of the $\tau$ integral.  The
parallel transport of the charge $(p_1^a {\cal J}_a + j_1^a {\cal P}_a)$
is along a straight path from $\bx_1$ to $\bx_1 + ({\bf a}_1 + {\bf a}_2)\tau$,
then along another straight path to $\bx_1 + {\bf a}_1 \tau + {\bf a}_2 \tau'$.

Following earlier approaches [4,5,6], a complete set
of physical observables may be constructed from the holonomies
of the ISO(2,1) gauge field around the locations of the
particles\note{\tenpoint\noindent Note that,
as a consequence of $\lbrace\phi_{\pi \, i}^A,x^j_B\rbrace \not=
0$, the particle locations $x^i_A$ are not physical observables.}.
Introduce two loops based at $\bx_*$ which encircle the particles as drawn in
Fig.~1 (any basis for the fundamental group will do), and define
$$\eqalign{
h_1=
\exp \left \lbrace -{1 \over 2} \left( \bar p_1^a {\cal J}_a + \bar\jmath_1^a
{\cal P}_a \right) \right\rbrace
&= {\rm P}\, {\rm exp} \left\lbrace - \oint\limits_{\hbox{\rm loop 1}}
\left(e^a_i {\cal P}_a + \omega^a_i{\cal J}_a\right) dx^i \right\rbrace\cr
h_2=
\exp\left\lbrace - {1 \over 2} \left( \bar p_2^a {\cal J}_a + \bar\jmath_2^a
{\cal P}_a  \right) \right \rbrace
&= {\rm P}\, {\rm exp} \left\lbrace - \oint\limits_{\hbox{\rm loop 2}}
\left(e^a_i {\cal P}_a + \omega^a_i{\cal J}_a\right) dx^i \right\rbrace.\cr}
$$
One may make $\bar p_A^a$ and $\bar\jmath_A^a$ into physical observables by
locating the base point $\bx_*$ at spatial infinity, where ISO(2,1) gauge
transformations drop off to the identity -- this makes $\bar p_A^a$,
$\bar\jmath_A^a$ ISO(2,1) invariant\note{\tenpoint\noindent Alternatively,
one can keep $\bx_*$ finite and form gauge invariant combinations of
$\bar p_A^a$, $\bar\jmath_A^a$ at $\bx_*$ by using the invariant forms on
the ISO(2,1) algebra.}.

Are the $\bar p_A^a$, $\bar\jmath_A^a$ invariant under the other constraints
as well?  Holonomies are left unchanged under the action of any single finite
transformation $U(a^i)$ since any choice of loops may be deformed so
that they do not intersect the string singularity generated by $U(a^i)$.
Consider however the effect of $B$ on the holonomy around the
second particle\note{\tenpoint\noindent The
transformation generated by $B$, although connected
to the identity, evidently cannot be written as a single exponential of a
generator.}.
{}From the definition of $B$ in (6) and the effect of $U(a^i)$ on the ISO(2,1)
gauge field in (5), it can be shown that the effect of $B$ is to break
loop 2 where it intersects the triangle defining $B$,
and to insert the group element associated with the charge
of particle 1 parallel transported counterclockwise around the triangle
to the intersection point.
If one considers only states which are annihilated by the ISO(2,1)
gauge constraints $\phi_R^a$ and $\phi_T^a$, then one may continuously deform
the path on which the holonomy is calculated.
The braid operator can thus be
seen to change the path so  that
$$
B \exp\Bigl\lbrace -{1 \over 2} \left( \bar p_2^a {\cal J}_a + \bar\jmath_2^a
{\cal P}_a \right) \Bigr\rbrace B^\dagger = \exp\Bigl\lbrace -{1 \over 2}
\left( \bar p'_2{}^a {\cal J}_a + \bar\jmath'_2{}^a
{\cal P}_a \right) \Bigr\rbrace
\eqno(7)
$$
(and similarly for $h_1$), where
the primed holonomies are calculated on the braided loops indicated in
Fig.~2.

Note that these braided loops may be obtained from the original loops by
conjugation with the total holonomy, i.e.~with loop 2 followed by loop 1.
This means that the primed charges at $\bx_*$ are obtained by a Poincar\'e
transformation,
$$
\bar p_A'{}^a {\cal J}_a + \bar\jmath_A'{}^a {\cal P}_a = e^{- {1 \over 2}
\left(P^a {\cal J}_a + J^a {\cal P}_a\right)}\left(\bar p_A^a {\cal J}_a +
\bar\jmath_A^a {\cal P}_a\right)e^{+ {1 \over 2}
\left(P^a {\cal J}_a + J^a {\cal P}_a\right)}
$$
where the total charges $P^a$, $J^a$ are defined by
$$
e^{-{1 \over 2} \left(P^a {\cal J}_a + J^a {\cal P}_a\right)} = e^{-{1 \over
2} \left(\bar p_1^a {\cal J}_a + \bar\jmath_1^a {\cal P}_a\right)}
e^{-{1 \over 2}\left(\bar p_2^a {\cal J}_a+\bar\jmath_2^a {\cal P}_a\right)}.
$$
Note also that the content of the braiding condition is independent of the
original choice of loops.  When more than two particles are present, there
will be an operator $B_{AB}$ defined as above corresponding to each pair
of particles.  From their construction these operators provide a
representation of the braid group [9].

We are now in a position to discuss a two-particle
wave functional that is invariant
under all of the constraints.  Any wave functional constructed using
$h_1$ and $h_2$, with an appropriate choice of
polarization, is guaranteed to solve the ISO(2,1) constraints.  As discussed
above, the local action of the $\phi^A_{\pi\,i}$ constraint
eliminates the dependance on $x_a^i$.  The
reparameterization constraint $\phi_{p^2}^A$ is easily expressed in terms of
the holonomies, since $\bar p^a_A$ and $\bar \jmath^a_A$ are related by
Poincar\'e transformations to $p^a_A$ and $j^a_A$ for each particle. As a
result, the constraint may be rewritten as $\phi^A_{\bar p^2}={\bar
p^2_A-m^2_A}.$ The effect of this constraint on the wave functional follows
from the Poisson algebra of the $\bar p^a_A$ and $\bar\jmath_A^a$ [10].
In order to complete the quantization, it remains only to demand invariance
under the braiding operator $B$.

Eq. (7) shows that the effect of $B$ is equivalent to the condition
imposed in Ref. 6 to implement invariance under the mapping class
group.  Note, however, that in our formulation the topology of the spatial
hypersurface is
taken to be ${\rr}^2$, so that all diffeomorphisms are connected to the
identity. The braiding condition arises from the action of the
$\phi^A_{\pi\, i}$ constraint which describes how spatial diffeomorphisms
change the positions of the particles on ${\rr}^2$.

A possible
choice of polarization is to take the wave functional
to be a function of the $\bar p^a_A$.
The functional satisfies all the constraints if $\bar p^2_A=m_A^2$,
and $B\Psi(\bar p^a_A)=\Psi(\bar p^a_A)$.
A more complete
discussion of the consequences of the braiding condition for two
particles, and some words
on the role of surface terms and boundary conditions, may be
found in Refs. 6 and 10.

\subhead{\bf Acknowledgements}

The authors thank Michael Crescimanno, Fay Dowker, Roman Jackiw,
David Kastor, Jennie Traschen and Erik Verlinde for
valuable discussions.

\subhead{\bf References}

\item{1.} S.~Deser, R.~Jackiw, and G.~'t Hooft, {\it Ann.~Phys.} {\bf
152}, 220 (1984); J.~R.~Gott and M.~Alpert, {\it Gen.~Rel.~Grav.} {\bf 16},
243 (1984); S.~Giddings, J.~Abbott, and K.~Kuchar, {\it Gen.~Rel.~Grav.}
{\bf 16}, 751 (1984).

\refis{2.} See S. Carlip, {\it Six ways to quantize (2+1)-dimensional
gravity}, Davis Preprint UCD--93--15, gr-qc/9305020, and references
therein.

\refis{3.} A.~Ach\'ucarro and P.~K.~Townsend, {\it Phys. Lett} {\bf 180}, 85
(1986).

\refis{4.} E.~Witten, {\it Nucl.~Phys.} {\bf B311}, 46 (1988).

\refis{5.} G.~'t Hooft, {\it Comm.~Math.~Phys.} {\bf
177}, 685 (1988); E.~Witten, {\it Nucl.~Phys.} {\bf B323}, 113 (1989);
Ph.~Gerbert, {\it Nucl. Phys.} {\bf B346}, 440 (1990); M.~K.~Falbo--Kenkel and
F.~Mansouri, {\it J.~Math.~Phys.} {\bf 34}, 139 (1993); G.~'t Hooft, {\it
Canonical Quantization of Gravitating Point Particles in 2+1 Dimensions},
Utrecht preprint, gr-qc/9305008 (May 1993).

\refis{6.} S.~Carlip, {\it Nucl.~Phys.}
{\bf B324}, 106 (1989).

\refis{7.} G.~Grignani and G.~Nardelli, {\it Phys.~Rev.} {\bf D45}, 2719
(1992).

\refis{8.} R.~Jackiw, in {\it Current Algebra and Anomalies},
eds.~S.~B.~Treiman, R.~Jackiw, B.~Zumino, and E.~Witten (Princeton, 1985),
pp.~311 -- 320.

\refis{9.} J. Birman, {\it Braids, Links and Mapping Class Groups.} Annals
of Mathematical Studies \# 82. Princeton University Press, Princeton, 1973.

\refis{10.} D. Kabat and M. E. Ortiz, in preparation.

\subhead{\bf Figures}

\noindent {\bf Figure 1.} The cocycle measures the flux through the
triangle, and the holonomies are taken around the indicated loops.
\medskip
\noindent {\bf Figure 2.} Loops obtained from Fig.~1 by the action of
the braid operator.

\vfill
\end